\documentclass{article}
\pdfoutput=1
\usepackage{times}
\usepackage{latexsym}
\usepackage{amsfonts,amssymb,amsmath}
\usepackage{alltt}
\usepackage{graphicx}
\usepackage{xspace}

\newcommand{\eg}[0]{e.g.,\xspace}
\newcommand{\etc}[0]{etc.\xspace}

\newtheorem{THEOREM}{Theorem}
\newenvironment{theorem}{\begin{THEOREM} \hspace{-.85em} {\bf :} }%
                        {\end{THEOREM}}
\newcommand{\prf}{\noindent{\bf Proof:} }
\newcommand{\eprf}{\bbox\vspace{0.1in}}
\newcommand{\bbox}{\vrule height7pt width4pt depth1pt}

\usepackage{named}
\renewcommand{\citeyear}{\shortcite}
 \newcommand{\fullv}[1]{#1}
 \newcommand{\shortv}[1]{}
\newcommand{\commentout}[1]{}

\title{Decentralised Norm Monitoring in Open Multi-Agent Systems}

\author{
Natasha Alechina\\
University of Nottingham\\
\texttt{nza@cs.nott.ac.uk}
\and
Joseph Y. Halpern\thanks{Supported in part by NSF grants IIS-0534064,
  IIS-0812045, 
IIS-0911036, and CCF-1214844, and by AFOSR grants
FA9550-08-1-0438, FA9550-09-1-0266, and FA9550-12-1-0040,
and ARO grant W911NF-09-1-0281.}\\
Cornell University \\
\texttt{halpern@cs.cornell.edu}
\and
Ian A. Kash\\
 Microsoft Research \\
\texttt{iankash@microsoft.com}
\and
Brian Logan\\
University of Nottingham \\
\texttt{bsl@cs.nott.ac.uk}
}

\date{}

\begin{document}
\maketitle

\begin{abstract}
We consider the problem of detecting norm violations in open multi-agent systems (MAS). We show how, using ideas from \emph{scrip systems}, we can design mechanisms where the agents comprising the MAS are incentivised to monitor the actions of other agents for norm violations.  The cost of providing the incentives is not borne by the MAS and does not come from fines charged for norm violations (fines may be impossible to levy in a system where agents are free to leave and rejoin again under a different identity). Instead, monitoring incentives come from (scrip) fees for accessing the services provided by the MAS. In some cases, perfect monitoring (and hence enforcement) can be achieved: no norms will be violated in equilibrium.  In other cases, we show that, while it is impossible to achieve perfect enforcement, we can get arbitrarily close; we can make the probability of a norm violation in equilibrium arbitrarily small. We show using simulations that our theoretical results hold for multi-agent systems with as few as 1000 agents---the system rapidly converges to the steady-state distribution of scrip tokens necessary to ensure monitoring and then remains close to the steady state.
\end{abstract}

\section{Introduction}

Norms have been widely proposed as a means of coordinating and
controlling the behaviour of agents in a multi-agent system
(MAS). Norms specify the behaviours that agents should follow 
to achieve the objectives of the MAS.
For example, the designer of a system to allow agents to post content (invitations to
tender for work, prices of goods or services, etc.) may wish to ensure
that the content posted is relevant, accurate and up to date. 

In a MAS where norms must be enforced, the responsibility for 
enforcing norms lies with a system component termed the \emph{normative organisation} \cite{Dastani//:09a}, which continuously monitors the
actions of the agents (and perhaps carries out other tasks on behalf of the MAS). 
If an action (or the state resulting from an action) would violate or
violates a norm, the action is either prevented, or the agent that
performed the action is penalised (incurs a sanction). The effective
monitoring of agent actions is therefore key to enforcing norms in a
MAS. However, in large systems with many agents, 
maintaining a separate component to monitor the actions of the agents
may involve significant overhead for the MAS.

In this paper, we propose an approach to norm monitoring in open
multi-agents systems in which the monitoring of agent actions
is performed by the agents comprising the MAS. We term
this \emph{decentralised monitoring}. 
We focus on norms which prohibit certain actions (or the resulting
state), for example, posting irrelevant or inaccurate content may be
prohibited. 
The novelty of our approach is that the MAS does not need to bear the cost of
paying for monitoring; at the same time we do not need to assume 
that fines can be levied on the agents who violate the norms and used to
pay for monitoring, as done by Fagundes et
al. \citeyear{Fagundes//:14a}. The latter assumption 
does not hold for many open systems 
where the agents can always leave the system and, if needed, rejoin it later
under a different identity. Hence, a key issue for our approach is how to
incentivise the agents to monitor the actions of other agents. 
We show how, using ideas from \emph{scrip
systems} \cite{Friedman//:06a}, we can design incentive-compatible
mechanisms where the agents do the monitoring themselves. 
We can think of scrip as ``virtual money'' or ``tokens''. 
Performing an action costs a token, and detecting violations is
rewarded with tokens. 
The main difference between our setting and that of \cite{Friedman//:06a} is
that the agents are not always rewarded after they monitor,
but only if they discover a violation. This requires a non-trivial adaptation
of the techniques developed in \cite{Friedman//:06a}. 

We consider two settings.  In the first, the \emph{inadvertent setting},
actions that violate a norm are assumed to be inadvertent or unintentional:
violating a norm does not increase an agent's utility. 
In the second, the \emph{strategic setting}, 
actions that violate the norm are intentional:
violating the norm increases the agent's utility, and 
an agent chooses whether to try to violate the norm.
We describe a mechanism that achieves \emph{perfect enforcement} in
the inadvertent setting; in equilibrium, all actions are
monitored and hence there are no violations of the norm.  In the
strategic setting, we prove that there can be no equilibrium with
perfect enforcement.  However, the probability of violations can be
made arbitrarily small: for all $\epsilon > 0$, we can design a
mechanism where, in equilibrium, the probability of violations is~$\epsilon$.  
We show how all the key steps in our mechanisms can be decentralised, 
and how our ideas can be extended to \emph{open}
systems, where agents may enter and leave the system at any time.  
We also consider robustness, and show that the mechanisms we propose
are \emph{$m$-resilient}: no coalition of up to $m$ agents can increase
their utility through collusion. 
Finally, we show using simulations that our theoretical results hold
for multi-agent systems with as few as 1000 agents. In particular, we
show that the system rapidly converges to the steady-state
distribution of scrip tokens necessary to ensure monitoring and then
remains close to the steady state.

\section{Incentivising Monitoring}

In this section we outline the simple scenario that we use as a running example throughout the remainder of the paper.  

We consider a MAS where agents want to post
content on the web. There are norms regarding what may be posted; for
example, copyrighted images should not be posted, and comments should
not be abusive or defamatory.  We assume that agents may occasionally
submit posts that violate the norm. If such content appears on the web, the
MAS loses significant utility (e.g., it can be fined or sued). 
Note that here we are viewing the MAS as a whole as an entity that can
be fined or sued for norm violations, 
and which may incur the computational costs associated with monitoring
for violations (and hence can gain and lose utility).   

It is therefore in the MAS's interest
that submitted posts are checked for compliance with the norm
before they appear on the web. We assume that it
is possible to check objectively if a particular item of content
violates the norm. (For simplicity, we assume that if a post that
violates the norm is checked, the violation will be detected.  We can
easily modify our approach to handle the case where there is some
probability $\rho$ of the violation being caught.) 
Checking whether a post is `bad' 
(violates the norm) requires some work, and incurs a small utility cost.
Although checking requires some resources, we assume that if a
violation is found, evidence of the violation can be provided that can
be checked in negligible time (so we do not need to deal with disputes about
whether  content violates the norm). If the content does violate the norm, 
the post is discarded and no violation occurs.
%
We assume a basic infrastructure that ensures that
content posted by an agent is signed, and that the digital signatures
can be trusted. The signature identifies the agent id, and is interpreted 
as a statement by the agent that the content posted conforms to the norm. 
Note however, that the infrastructure does not itself enforce
the norm; it serves only to ensure auditability. We believe that such a
separation of concerns is good design: the same basic infrastructure
may be used by different systems with different norms. 

There is a system-level objective that content conform to the norm,
but since the cost to the MAS to check all posts may be prohibitive, 
we would like to distribute the monitoring of posts among
the agents that use the system.  Just as for the MAS, 
monitoring incurs a small negative utility for an agent.  This means that 
agents must be appropriately incentivised to monitor.
It should be clear that the ideas exemplified by this scenario are applicable far more broadly.

We formalise the posting and monitoring of content for norm violations
as a non-cooperative game.  This scenario (and the resulting game) is similar to the 
scenario considered by Friedman et al.~\citeyear{Friedman//:06a}, but
differs in several 
key respects.  Friedman et al.~assume that some agent
requests a service (e.g., downloading a file), and the problem is to
incentivise provision of the service; if the service is not provided,
the requesting agent will not be satisfied. 
Here, it is not necessary that each post be monitored for the posting
agent to be satisfied. We assume that, if no agent monitors, it is
possible for the posting agent to post and benefit from it; however, a
norm violation may be missed.  
\fullv{This difference turns out to be not so significant.}
A more significant difference is that, in our setting, a post may violate
the norm.  This has no analogue in the setting of Friedman et al., and
does complicate matters, as we shall see. 
Despite this, many of the ideas used by Friedman et
al. \citeyear{Friedman//:06a} can be used in our setting.  In
particular, we adopt the idea of using \emph{tokens} as
payment for posting and as a reward for monitoring. In order to post,
an agent must pay one token; finding a bad post is rewarded by
receiving one or more tokens as payment. This encourages agents to volunteer to
monitor posts.  The exact mechanisms and amounts are
discussed below.   

We consider two scenarios, one in which bad posts are unintentional, and one in which they are strategic.
\fullv{These correspond to different games.}
We formalise these two scenarios 
\fullv{and our approach for dealing with them}
in the next two subsections.

\subsection{Unintentional Violation} \label{sec:unintentional-violation}

In this scenario, bad posts happen with a constant probability
$b$, but agents are unaware that they are violating the norm when they
post something inappropriate.
For technical reasons, we assume that $b$ is a rational number
(our results hold as long as we use a
sufficiently good approximation to the true probability, so this
assumption is really without loss of generality).
The game in the inadvertent scenario is described by the following parameters:
\begin{itemize} \setlength{\itemsep}{-1pt}
\item a finite set of $n$ agents $1, \ldots, n$;

\item the time between rounds is $1/n$;%
\footnote{The assumption that the time between rounds is $1/n$, which
also made by Friedman et al. \citeyear{Friedman//:06a}, makes the  analysis easier. 
It guarantees that, on average, each agent wants to post one message
per time unit, independent of the total number of agents.}

\item at each round $t$ an agent is picked at random 
to submit a post 
(we implicitly assume that agents always have something that
they want to post);

\item probability of a post being bad: $b$;

\item utility of posting (to the agent doing the posting): $1$
(independent of whether what is posted violates the norm);

\item disutility of monitoring (to the agent doing the monitoring): 
$-\alpha$ (where $0 < \alpha < 1$);

\item discount rate: $\delta \in (0,1)$.
\end{itemize}
The game runs forever.    As is
standard in the literature, we assume that agents discount future
payoffs. This captures the 
intuition that a util now is worth more than a util tomorrow, and
allows us to compute the total utility derived by an agent in the
infinite game.
We have assumed for simplicity that the system is homogeneous: all
agents get the same utility for posting (1), the same disutility for
monitoring ($-\alpha$), have the same probability of being chosen to post
something ($1/n$), and have the same discount factor ($\delta$).
Using ideas from
\cite{Kash//:12a}, we can extend the approach discussed here
to deal with different \emph{types} of agents, characterised by
different parameters.  Some agents may want to post more often; other
agents may be less patient (so have a smaller discount rate); \etc 

We need some additional notation to describe what happens:
\begin{itemize}
\item $p^t \in \{1, \ldots, n\}$ is the agent chosen to submit a post in round $t$;

\item $v^t \in \{0, \ldots, n\} \setminus \{p^t\}$;  $v^t = j$ if agent $j \neq p^t$ is chosen to
monitor in round $t$, and $v^t = 0$ if no one is chosen to monitor at
round $t$; 

\item $f^t \in \{0,1\}$; $f^t = 0$ if the content posted in round
$t$ is good, $f^t = 1$ if it is bad.
\end{itemize}

Given that good and bad posts have the same utility (1), the
utility of an agent $i$ in round $t$ is: 
\[ u_{i}^{t} = \left\{
\begin{array}{ll}
1 & \mbox{if $i = p^t$ and either $v^t=0$ or $f^t = 0$;}\\
-\alpha & \mbox{if $v^t = i$;}\\
0 & \mbox{otherwise.}
\end{array}
\right . \]
Thus, an agent gets utility 1 at round $t$ if it is chosen to submit a
post ($p^t=i$), and either the post is not monitored ($v^t = 0$) or it does not
violate the norm ($f^t = 0$).
Given the discount factor $\delta$,  the total utility $U_i$ for agent
$i$ is $\Sigma^{\infty}_{t=0} \delta^{t/n}\ u^t_i$.

So now the question is how to incentivise monitoring.  The key idea is
to use tokens as payment for posting and as a reward for
monitoring. Note that the number of tokens that an agent has does not
affect  the agent's utility.  However, if an agent requires a token
to post something, the number of tokens that an agent has does have an
indirect impact on utility; if the agent has no tokens, then it will
not be able to post anything, and thus will forego the opportunity to
get utility 1. 

Agents are rewarded with tokens only if they detect a bad post.   
We argue below that in order for the system to function successfully (agents
being able to post, and some agents always available for monitoring), the
`right' amount to pay for finding a bad posting is $1/b$.%
\footnote{We are implicitly assuming that tokens are divisible into
units such that it is possible to transfer $1/b$ tokens.} 
This means, in expectation, an agent gets one token for finding a bad posting.  
Thus, the price of a posting is equal to the expected cost of checking a posting.  

Paying agents $1/b$ for finding a bad posting 
makes the situation similar to that in \cite{Friedman//:06a}, where
the agent wanting work done pays one token, and the agent
doing the work gets one token. However, the fact that in the
current setting payment is made only if a problem is found 
complicates matters. An expected payment of $1$ token  is not equivalent
to an actual payment of $1$ token!
To understand the issue here, note that
the most obvious way to deal with the payment of tokens is to have the
agent who wants to post pay one token to the 
normative organisation, and then have the normative organisation pay
$1/b$ tokens to the monitor if a violation is detected. 
But there are problems with this approach.  If
monitors have a long run of ``bad luck'' and do not find postings that
violate the norm, there will be very few tokens left in the system;
on the other hand, if monitors get lucky, and find quite a few
postings that violate the norm, the normative organisation will end up 
pumping quite a few tokens into the system.  As pointed out by Friedman et
al. \citeyear{Friedman//:06a}, having both too few or too many
tokens in the system will cause problems. Intuitively,
with too many tokens in the system, (almost) everyone will have plenty of
tokens, so no one will volunteer to monitor; with too few tokens in the system,
it will often be the case that the agent who wants 
to post will not have a token to pay for it.%
\footnote{The situation would be even worse if the payment
for detecting a violation were different from $1/b$; then after some
time there would certainly be too few or too many tokens in the
system.}   This problem  
does not occur in the setting of Friedman et
al.~\citeyear{Friedman//:06a}, because there, the payment of person
doing the work always matched exactly the payment received by the
person doing the work.  

We deal with this problem by having the agents rather than the normative organisation perform the role of the ``bank''.  
When agent $i$ wants to post, it pays a randomly chosen agent who has fewer than the maximum number of tokens allowed (see below) 1 token; if an agent $j$ monitors  and finds a violation, a randomly chosen agent with at least $1/b$ tokens gives $j$ $1/b$ tokens.
This ensures that the number of tokens in 'circulation' remains constant.
(Note also that a randomly chosen agent pays the monitoring agent
$1/b$ tokens if there is a violation; this allows agents to post as
long as they have a single token.)

We assume that all agents follow a \emph{threshold strategy} when deciding whether to monitor.  There is a fixed threshold $k$ such that agents volunteer iff they have fewer than $k$ tokens. It is easy to see that there is an equilibrium in threshold strategies if everyone uses a threshold of 0.  In that case, no one ever volunteers to monitor a posting, so everyone gets to post, without monitoring.  Of course, no agent has any incentive to deviate from this strategy.  On the other hand, this equilibrium is rather bad from the point of view of the MAS.  We are thus interested in nontrivial equilibria in threshold strategies, where everyone uses a threshold $k > 0$.  

Friedman et al.~\citeyear{Friedman//:06a} show that, in their setting, there is a
nontrivial equilibrium in threshold strategies; more precisely, for all $\epsilon > 0$, there exists a $\delta$ sufficiently close  to 1 and a threshold $k$ such that as long as the discount factor is at least $\delta$, all agents using a threshold of $k$ is an $\epsilon$-Nash equilibrium: no agent can gain more than $\epsilon$ by deviating.

We can get a similar result in our setting, using the banking idea above, where the maximum number of tokens any agent may have is $k + 1/b$. To summarise, if an agent $i$ has at least one token and is chosen to submit a post ($i = p^t$),  $p^t$ gives a randomly chosen agent with fewer than $k + 1/b$ tokens one more. The posting agent $p^t$ than asks for volunteers to act as monitor.  All agents with fewer than $k$ tokens volunteer. If at least one agent volunteers, one, $v^t$, is chosen at random to act as monitor. If $v^t$ confirms the post conforms to the norm, it is posted.  If $v^t$ detects a
violation of the norm, then the post is discarded, and a randomly chosen agent with at least $1/b$ tokens gives $v^t$ $1/b$ tokens.

\begin{theorem}\label{thm1} For all $\epsilon > 0$, there exist a
$\delta$ sufficiently close to 1 and an $n$ sufficiently large such that if all 
$n$ agents have a discount factor $\delta' \ge \delta$, then there exists
a $k$ such that the mechanism above with all agents using a
threshold of $k$ is an $\epsilon$-Nash equilibrium. 
\end{theorem}

\fullv{
The proof of Theorem~\ref{thm1} is similar in spirit to that of
Friedman et al.~\citeyear{Friedman//:06a}. A proof sketch can be found in Section \ref{sec:proof}.
}
%

Note that in the equilibrium the existence of which is stated in Theorem~\ref{thm1}, we get perfect enforcement; all bad posts will be detected.
Although the theorem applies only if $\delta$ is ``sufficiently close
to 1'' and $n$ is ``sufficiently large'', simulations (reported in 
Section~\ref{simulations}) show that, in practice, the distribution of tokens 
is reasonably stable with $n$ as small as 1000.%
\footnote{In fact, simulations show that
  even with 100--1000 agents, performance is in a range that seems
  quite tolerable in practice.}
These simulations also show that, rather than having one randomly
chosen agent pay $1/b$ tokens if a violation is discovered, we can
have $\lceil 1/b \rceil$ randomly chosen agents pay 1 token each. 
The latter approach may be more acceptable in some systems.

\subsection{Strategic Violation}

We now consider the scenario of strategic violation. 
In this scenario, we assume when an agent is chosen to 
submit a post, it can either submit something good (i.e., that does
not violate the norm) or something bad. 
The parameters of the game are the same as in
Section \ref{sec:unintentional-violation}, except that there is no
longer a probability $b$ of a posting being bad (the quality of
a posting now becomes a strategic decision), and the utility of a bad
posting is no longer 1, but $\kappa > 1$. (We must assume $\kappa > 1$
here, otherwise no agent would ever post anything bad:
the utility of doing so is no higher than that of posting something
good, and the violation may be detected.)

As before, monitoring agents get paid only if they find a bad post.
With these assumptions, it is not hard to show that there does not exist an
equilibrium with perfect enforcement.

\begin{theorem} In the setting of strategic violations, there can be no
equilibrium with perfect enforcement.
\end{theorem}

\prf Suppose, by way of contradiction, that there is an equilibrium
with perfect enforcement. 
In this equilibrium, all attempts to make a bad posting are
caught.  Thus, no agent will use a strategy that gives a positive
probability to making a bad posting, for that agent would get higher
utility by posting something good instead of something bad.  But then
no agent would 
monitor.  Even if an agent volunteers to monitor, 
the agent would just claim that no violations were found
without actually monitoring (since monitoring costs $-\alpha$ in
utility, and there would be no 
violations to catch).   But if there is no actual monitoring, then
agents should deviate and make bad postings, since they will not be
caught.  \eprf

Although we cannot achieve perfect enforcement in the strategic
setting, we can achieve the next best thing: we can
make the probability of a bad posting as low as we want.  More
precisely, for all
$\epsilon, \epsilon' > 0$, there is an $\epsilon$-Nash equilibrium such
that the probability of a bad post is $\epsilon'$.  

The idea now is that, with some probability, a 
submitted post will not be
checked; there will be no attempt to get volunteers to monitor that
posting.  
Let $c^t =0$ if there is no monitoring in round $t$; $c^t = 1$ otherwise.  
The decision regarding whether to monitor is made \emph{after} the
poster submits their post (otherwise, the agent chosen to
post will always post something bad in round $t$ if $c^t = 0$).  
If $c^t = 0$, then whatever the poster 
submits is posted in that round, whether it is good or bad.
As before, if an agent submits a  bad post and there is
monitoring,  
we assume that the bad post definitely will be detected and discarded, so the
posting agent gets utility 0 in that round.
The utility of agent $i$ in round $t$ now becomes 
\[ u_{i}^{t} = \left\{
\begin{array}{ll}
1 & \mbox{if $i = p^t$ and $f^t=0$};\\

\kappa & \mbox{if $i = p^t$, $f^t=1$, and either $c^t =0$}\\
&\mbox{or $c^t = 1$ and $v^t = 0$}; \\

-\alpha & \mbox{if $i = v^t$};\\

0 & \mbox{otherwise.}
\end{array}
\right . \]

Suppose that the normative organisation decides  that
postings will be monitored with probability $1 - 1/\kappa$.  
Further suppose that an agent uses a randomised algorithm: 
with probability $\beta$ it 
submits a good posting, and with probability $1-\beta$ it submits a bad posting.
\fullv{Note that the} \shortv{The} agent's expected payoff is then 
$\beta + (1-\beta)(1/\kappa)\kappa = 1$, independent of $\beta$.
Thus, we get an equilibrium in the single-shot game if monitoring occurs with
probability $1- 1/\kappa$ and agents submit bad postings with
probability $\beta$, for all choices of $\beta$, provided that there
is always guaranteed to be a monitor available.  We will show that
again there is an equilibrium in threshold policies. 
As long as there are not too many tokens in the system, there are bound 
to be  some agents with fewer than the threshold number of tokens, so
there will  be a volunteer.

We assume the designer of the MAS specifies a value
$\beta^*$ (which, intuitively, should be a small `tolerable'
probability of a violation occurring).  
If a monitor that finds a problem is paid $1/\beta^*$
tokens, then essentially the same type of mechanism as that proposed for
the case of 
unintentional violations will work, provided that we get bad
postings with probability exactly $\beta^*$.  So, perversely, in this
setup, while all strategies are equally good for the poster, the
MAS actually wants to 
\emph{encourage} agents to post something bad with probability $\beta^*$,
so that monitors again get an expected payment of 1 token.
The way to do this is for the normative organisation to announce
that it will track the number of bad postings, and if the fraction of
postings that have been bad up to round $t$ 
is $\beta$, checks will happen with probability $1
- \beta^*/(\beta \kappa)$.  Thus, if  $\beta=\beta^*$,
then checks happen with probability $1-1/\kappa$, and we have an
equilibrium.  Moreover, the payment ($1/\beta$ tokens) is exactly what
is needed to ensure that, in equilibrium, monitoring occurs with
probability $\beta^*$.  For if
$\beta < \beta^*$, then the check will happen with probability less than $1 -
1/\kappa$, which means that agents will want to make more bad posts.  On the
other hand, if $\beta > 
\beta^*$, then monitoring will happen with probability greater than $ 1 -
1/\kappa$, and agents will want to make fewer bad posts.  Thus, in
equilibrium, we get bad posts with probability exactly $\beta^*$.  

To summarise, we have the following mechanism, given a threshold $k$.
If an agent is chosen to post, it submits bad content with
probability $\beta^*$ and good content with probability $1-\beta^*$.
After the agent has decided what to post and made the posting
available, the normative organisation decides whether the posting will be
monitored.  For an initial period (say 1,000 rounds), a posting is
monitored with probability $1 - 1/\kappa$; afterwards, if the fraction
of postings that have been discovered to be bad due to monitoring is
$\beta$ and $\beta$ is more than (say) two standard deviations 
from $\beta^*$, then monitoring occurs with probability $1
- \beta^*/(\beta \kappa)$.
If the decision has been made to monitor, and the posting agent
has at least one token (so that a post can be made), the posting agent
asks for volunteers and 
all agents with fewer than $k$ tokens volunteer to monitor and 
one is chosen to be the monitor. 
As in the case of unintentional violations, if at least one agent
volunteers, then the posting agent gives a randomly chosen agent with less
than $k + 1/\beta^*$ tokens one more.  If the monitor approves
the posting, it is posted.  If the monitor finds a problem with the posting, 
then a randomly chosen agent with at least $1/\beta^*$ tokens
gives the monitor $1/\beta^*$ tokens.

\begin{theorem}\label{thm2} 
For all $\epsilon > 0$, there exist a
$\delta$ sufficiently close to 1 and an $n$ sufficiently large such that if all
$n$ agents use a discount factor $\delta' \ge \delta$, then 
there exists a $k$ such that the mechanism above with all agents using a
threshold of $k$ is an $\epsilon$-Nash equilibrium. 
\end{theorem}
\shortv{
A proof sketch can be found at the anonymous url https://sites.google. com/site/aamas2016paper420/proofs.pdf.}
A proof sketch can be found in Section \ref{sec:proof}.

Note that, in the equilibrium whose existence is stated in
Theorem \ref{thm2}, 
the probability of a bad posting is $\beta^*$, as desired.

\subsection{Optimising Social Welfare}\label{sec:socialwelfare}

In this section, we expand our analysis to include the utility of the 
MAS, and show how social welfare can be maximised
by controlling the  number of tokens in the system.

Suppose that the MAS gets utility $-C$ (where $C > 0$) for each
norm violation.  In the setting where bad posts are inadvertent, if there
is no monitoring, the MAS suffers an expected loss
of utility of $bC$ in each round and the remaining players get 1 unit
of utility in each round (because a post is never discarded), 
so all the agents in the system get an
expected utility of $1-bC$ per round.  With monitoring, if we assume
for simplicity that an agent always has a token
when it wants to post something
and there is always a volunteer to monitor, players get a
total expected utility of $1-b - \alpha$ per round.
Thus, as long as $C > (b+\alpha)/b$, 
we maximise social welfare by monitoring.    
This analysis doesn't change if occasionally an agent
does not have a token to pay for a posting.

In the strategic setting, without monitoring, players will always post
inappropriate material, so the total utility will be $\kappa - C$ per
round.%
\footnote{Of course, the utility of an inappropriate post may decrease
if everyone is posting such material.}  With monitoring, assuming
agents post something 
bad with probability $\beta^*$, as we have seen,
the expected utility of an agent who makes a posting is 1 (independent
of $\beta^*$, so the total expected utility in each round is
$1 - (1-1/\kappa)\alpha$.  So monitoring increases social welfare 
if $C > \kappa - 1 + (1-1/\kappa)\alpha$.

Although we have taken the MAS to be a separate
entity with its own utility, in some cases the utility of the MAS
is best thought of as shared among the individual agents posting content
(e.g., in the 
case of a community or collaboratively maintained website).  In this 
case, we can assume that the players collectively bear the costs of a bad
posting.  It is then certainly reasonable to assume that
monitoring increases social welfare; otherwise the players would
simply not bother.  But even if the MAS is really an
independent entity, \fullv{whether or not monitoring increases social
welfare,} the MAS can ensure the monitoring equilibrium is the one that
occurs by simply posting this equilibrium and asking agents to play
it.  As long as sufficiently many agents play it (where ``sufficiently
many'' means that there are enough to ensure that there will always be
a monitor), then no agent gains by deviating.  Posting the equilibrium
is also useful if new agents join the system 
(see Section~\ref{sec:open}).  Note that
the MAS  can always threaten to shut down  if there is no
monitoring (which would be in its best interests).  Given that,
\fullv{in the remainder of the discussion,}
we assume that agents play an equilibrium
with monitoring.
  
Social welfare is then maximised if, every time an agent is chosen to 
submit a post, it has a token to pay for posting and there is 
an agent who is willing to volunteer to act as a monitor.  
Friedman et al.~\citeyear{Friedman//:06a} show that, in their setting, social
welfare depends completely on the average number of tokens per agent.
Social welfare increases monotonically as the average number of tokens
per agent increases, until it reaches a critical point.
The key point is that the threshold used in
equilibrium \emph{decreases} as the average number of tokens per
agent increases.  The critical point  is the one where
the average number of tokens per agent is equal to the
threshold.  
At this point, no one is willing to volunteer, so social
welfare drops immediately to 0.  

Essentially the same arguments apply in our setting, although we need
to be a little careful.  In the setting of Friedman et al., after an
initial period, no agent has more than the threshold number of tokens
(since once they hit the threshold, they stop volunteering).  In our
setting, since an agent can receive $1/b$ tokens (or $1/\beta^*$ in the case
of strategic violations) in one
round by discovering a violation, the maximum number of tokens
than an agent can have is not $k$, but $k +
1/b$ (or $k + 1/\beta^*$), where $k$ is the threshold.
As long as the average number of tokens per agent is less
than  $k - (k-1)/n$, then there is always guaranteed to be a
volunteer.  Thus, social welfare is maximized if the average
number of tokens per agent, $a$, is as large as possible, while still being
less than $k - (k-1)/n$, where $k$ is the equilibrium threshold
corresponding to $a$, since a higher average  increases
the likelihood that an agent who is chosen to post can make a posting.

\subsection{Minimising the Role of the Normative Organisation}

Although, using our mechanisms, the normative organisation no longer
has to monitor postings, it still has a role to play.  The normative
organisation: 
\begin{itemize}\setlength{\itemsep}{-1pt}
\item keeps track of the agents in the system (this is needed to ensure
that all agents are aware of a call for volunteers);

\item chooses an agent at random with fewer than $k + 1/b$ (or $k +
  1/\beta^*$) tokens to receive a token from the agent posting; 

\item choses an agent at random with at least $1/b$ (or $1/\beta^*$) tokens
to pay the monitor if the monitor detects a bad posting; 

\item keeps track of the number of bad postings and 
decides whether checking should be carried out in a given
round in the mechanism for strategic violations.
\end{itemize}
There is actually no need for the normative organisation to do any
of these things; we can distribute its role among the agents in the
system.  It is easy for the agents to maintain a list of the agents in the system
(think of this as a large email list).  Of course, it will have to be
updated whenever an agent enters or leaves the system 
(see also Section \ref{sec:open}).  

Choosing an agent at random from among a group of agents 
to receive a token from a posting agent, to monitor, or pay for
finding a bad post  
can be done in an incentive-compatible way (i.e., in a way that no agent has
any incentive to deviate) using the leader-election algorithm of 
Abraham et al.~\citeyear{Abraham//:13a}. Choosing a
leader among a set of players is equivalent to choosing one 
monitor among a set of volunteers.   In equilibrium, the algorithm
of \cite{Abraham//:13a} results in each player having an 
equal chance of being chosen. 

To handle the banking process, we can assume that each agent
keeps track of how many tokens each agent has.  All the transactions
can be announced publicly (i.e., who is chosen at random to make a
posting, who is chosen at random to get one token, etc.), so
everyone can update the amounts appropriately.  (We can minimise the
communication required by having only a small subset of agents keep
track of how many tokens each agent has, or by distributing the role of
the bank, so that each agent keeps track of the amounts held by only a
few other agents~\cite{Vishnumurthy//:03a}, but the overhead seems low in any case.)  

The leader-election algorithm of Abraham et al.~can 
also be used to decide whether checking should
be carried out in a given round in the second mechanism.  To see how this
works, first suppose that $\kappa$ is an integer that is at most $n$.
Choose a subset of $\kappa$ agents, including agent 1.
There is monitoring if the leader chosen among the $\kappa$ players is not
agent 1.  This guarantees that checking is done with probability $1-
1/\kappa$.  We can easily
modify this algorithm to compute any rational probability.
If the detection of a bad post is announced publicly,  
agents can also keep track of the number of bad postings in the 
mechanism for strategic violation case, 
so that the probability of checking can be modified
if needed.


\subsection{Open Systems}\label{sec:open}

We assume that our  system is open; agents can enter and leave at any
time.  Dealing with agents leaving is straightforward: they are just
deleted from the list of players.  Noticing that an agent has left is
also straightforward: an agent who does not attempt to make a posting
when it is chosen to submit a post, or does not pay tokens to a monitor 
that detects a bad post when it is chosen to, 
will be assumed to have left.  In the latter case, a new agent can be
chosen to make a payment by rerunning the leader election algorithm.

Dealing with agents entering is almost as straightforward.  We assume
that there is a url where agents can post a message saying that they
wish to join.  They are then automatically added (by all agents) to the
list of members.  As in Friedman et al.~\citeyear{Friedman//:06a}, we assume that a 
new agent starts out with no tokens, so new entrants
to the system cannot post anything.  This prevents an agent from joining
the system, not doing any monitoring until it runs out of 
tokens, then
leaving the system and rejoining under a different identity.  
New entrants can acquire tokens by monitoring, or by receiving one token
at random from an agent making a post.  

\newcommand{\factor}{F}
There is one other issue that needs to be dealt with: if the number of
agents changes significantly, the average number of tokens per agent
will change. As noted in
Section~\ref{sec:socialwelfare}, this can
have an effect on social welfare. To deal with this, we apply an idea
suggested by Friedman et al.: we choose a factor $\factor$, and multiply
the number of tokens that each agent has by $\factor$.  The factor
$\factor$ is chosen so as to make the average 
number of tokens close to
the threshold again.%
\footnote{If the desired average number of tokens is $t$, an obvious
alternative might be to have each agent entering the system start with
$t$ tokens.  But, as pointed out by Friedman et al., this leaves the
system open to a \emph{sybil attack}: a new player enters the system,
makes posts, but never monitors.   Once the agent runs out of tokens,
it drops out of the system and re-enters the system using a different
id.  This approach may be less problematic in our setting if agents
like to have their identity associated with a post, so there is some
loss of utility in leaving the system and re-entering with a new id.}
As long as all the agents are keeping track of the number of agents in
the system (or can consult a public url where this information is
available) and we are using public banking, then the agents can detect
if the average number of tokens has deviated significantly from the
target range, and apply the factor $\factor$ in a distributed way.

\subsection{Robustness}\label{sec:coalitions}

In a Nash equilibrium, no single agent can do better by deviating.
However, a coalition may be able to do better by deviating in a
coordinated way.  In large systems, it seems quite likely that
coalitions may form.  

Abraham et al.~\citeyear{ADGH06} define a strategy profile to be 
\emph{$m$-resilient} if no group of up to $m$ agents can increase
their utility by deviating. A Nash equilibrium is just a 1-resilient profile.
It is not hard to show that our basic protocol, combined with ``public
banking'', where all agents keep track of how many tokens are held by
each agent, is $m$-resilient 
\commentout{as long as $m$ is fewer than one-third
of the total number of agents. Essentially the same argument
as given earlier applies to groups, almost without change.
As long as there is no coalition that has at least one-third of the
agents in the system, we can use  \emph{Byzantine agreement} \cite{Fisbyz} 
to ensure that all agents agree on how many tokens each agent has.
}
for all $m$.  We assume that agents all keep track of
of the number of tokens that each agent has, and compare notes at each
step.  (If the number of messages exchanged is a concern, 
it should suffice to compare notes far less often.)  If there is 
disagreement between any pair of agents, then the system simply
stops.  Clearly, no group of agents gains any advantage by
misrepresenting the status of the bank.

We suggested that the leader-election algorithm of 
Abraham et al.~\citeyear{Abraham//:13a} could be used to distribute
the process of choosing a volunteer at random.  This algorithm is
 $m$-resilient for all $m < n$ in a completely-connected network
(which we have implicitly assumed---every agent can
communicate with every other agent). Thus, we can put all the pieces
together and still get $m$-resilience for reasonable-sized $m$.

However, this resilience claim relies on the assumption
that the actions and agents are exactly those in our model.
In the remainder of this section we discuss some alternatives
and our intuitions for how they would affect our results, but we
leave a full analysis to future work.

As pointed out by Friedman et al.~\citeyear{Friedman//:06a}, if it is
possible to 
transfer funds between agents, yet another type of collusion is
possible: coalitions can use a  lower threshold because they can
``insure'' each other  (i.e., if  one agent runs out of tokens to pay
for a posting,  another agent in  the coalition can loan it a
token); this allows  agents in the coalition to deviate by monitoring
less frequently.   However, this possibility does not arise in 
our setting with public banking: there is no procedure for one agent
to loan a token to another agent.  

If $b$ is relatively small (so that $1/b$ is large) and an agent $i$ could
somehow arrange that another  $j$ would always be the one 
to check $i$'s post (and vice versa), 
then $i$ might be tempted to deliberately make a bad posting, so
that $j$ could find it and collect $1/b$.
This is not a problem for Nash equilibrium, since it requires
collusion on the part of two agents, but it is an issue if we want to
prove 2-resilience.  
However, as our analysis
shows, in equilibrium, there will be many volunteers, so as long as
the choice among volunteers and the choice of agent required to pay if a
violation is found are both made at random, it is not hard to show
that type of deviation will not result in a gain, no matter how large 
the coalition is, since probability of gaining $1/b$ tokens
by deliberately deviating is equal to the probability of having to pay
$1/b$ tokens.

Another potential source of actions outside our model is asynchrony.  This
would complicate our Nash equilibrium analysis, but we would not expect
this to have a significant effect on single-agent deviations.  
With coalitions however, more care is needed.  In particular,
using the leader-election algorithm of Abraham et
al.~\citeyear{Abraham//:13a} is only $m$ resilient when $m < n/2$ with
asynchronous communication.

In practice, as pointed out by Abraham et al.~\citeyear{ADGH06}, we
may want even more.  We may want to allow for a certain number of
``irrational'' or ``malicious'' players, who do not to seem to be
acting according to 
their self-interests.  This may be simply because we do not understand
what motivates them; that is, we do not know their true utility
functions.  It may also be because of computer or system problems, or
unfamiliarity with the system; these are agents that would act rationally if
they could, but something is preventing them from doing so.  
Note that our proposal for dealing with the banking system,  
while sufficient to guarantee $m$-resilience, is
not even what Abraham et al.~\citeyear{ADGH06} call \emph{1-immune}:
one agent who wants to bring down the system can easily do so by lying
about the status of the bank.  We can deal with such ``malicious''
agents by using techniqes of \emph{Byzantine agreement} \cite{Fisbyz}
to get agreement on the bank status, as long as fewer than one-third
of the agents are malicious.
\fullv{Indeed, if there is a public-key infrastruture,
so that cryptography is available, we can actually use techniques of
Byzantine agreement to handle an arbitrary
number of malicious agents.}

While we think that other aspects of our system are quite robust, and
will degrade 
gracefully in the presence of such irrational or malicious behaviour,
we do not have a formal proof.  This is a 
topic that we believe deserves further exploration, since robustness
is an important property in practice.  

\section{Simulations}\label{simulations}

In this section, we use simulations to quantitatively evaluate the reasonableness
of our theoretical results.  In particular, we show that 
the system is well behaved
with as few as 1000 agents, converging to the steady-state
distribution of tokens quickly and then staying close to it.  
We also examine an alternative version of our 
system where we take the payment of $1/b$ tokens from $1/b$ random agents
rather than all from a single agent.

We report here results of simulations where there are twice as many tokens
as agents (so the average number of tokens per agent---shown to be a
key parameter in \cite{Friedman//:06a}---is two), agents used a threshold 
of $k = 5$, and, in the inadvertent setting, the  probability of a bad
posting is $b = 0.2$. (These choices are arbitrary; similar 
results hold for other settings of the parameters.) Results below refer
to the notion of closeness of distributions of tokens. We represent
each distribution as a vector that indicates the fraction of agents with
each amount of tokens and then calculate the Euclidean distance
between those vectors. 
We elected to use Euclidean distance to be consistent with prior
work~\cite{Friedman//:06a}. The claims below remain true for a variety of
reasonable notions of  
closeness of distributions.

\begin{figure}[htb]
\centering
\includegraphics[width=0.6\textwidth]{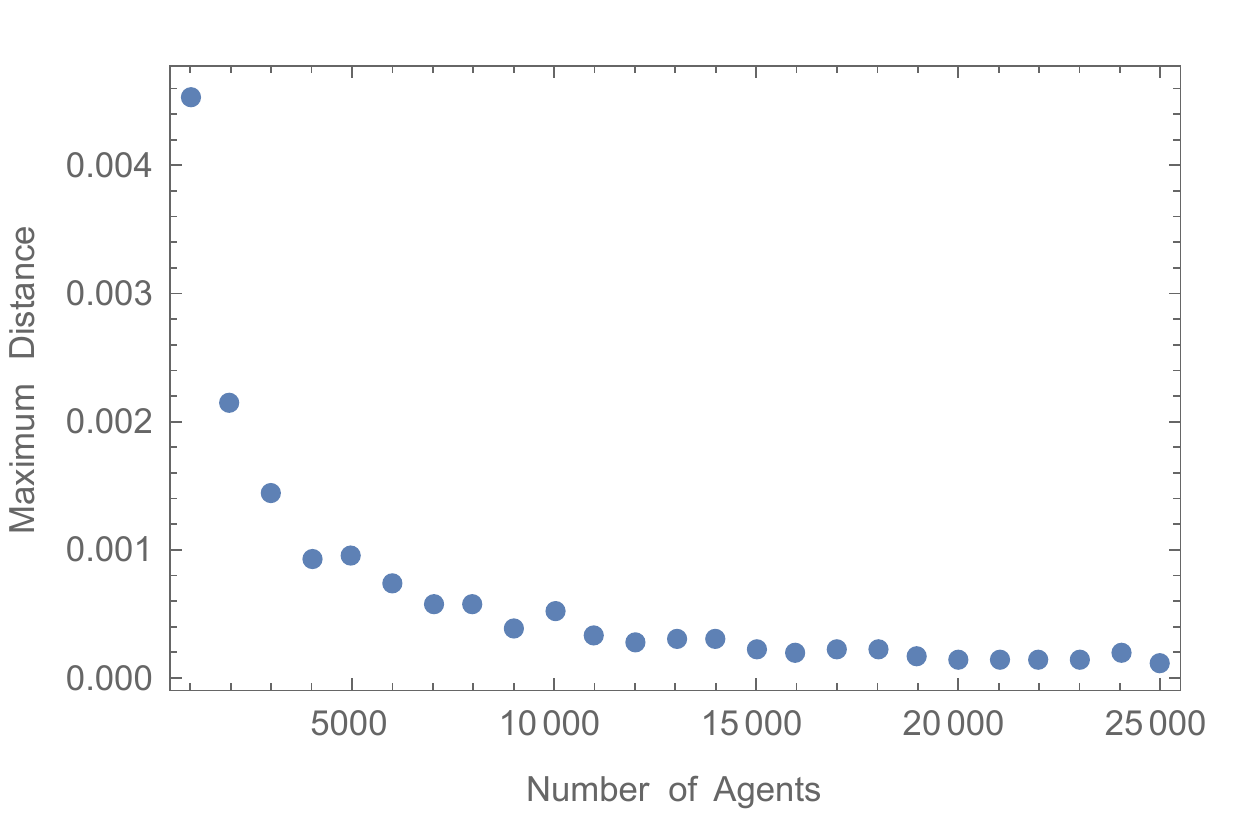}
\caption{The system stays close to the steady-state distribution of tokens.}
\label{fig:close}
\end{figure}

Figure~\ref{fig:close} shows the results of starting the system near the steady-state
distribution of tokens predicted by the theory and then running the system for 1 million
rounds.  It shows that the system stays quite close to the steady-state
distribution with as few as 1000 agents. For larger numbers, the system is even closer.
With larger numbers of agents, this simulation results in fewer rounds per 
agent, but
an alternate version where we ran it for 1000 rounds per agent (omitted) produced
visually indistinguishable results.

For 100 agents (omitted), the maximal distance is 10 times larger than in
Figure~\ref{fig:close} (below 0.05 rather than below 0.005), which may still 
be tolerable in practice. 

\fullv{In practice, it may be more natural to start the system with some
 more convenient distribution,  
 such as every agent having the same number of tokens, rather than the
 maximum-entropy distribution predicted by the theory.}
\fullv{(See Section~\ref{sec:proof}).}

\begin{figure}[htb]
\centering
\includegraphics[width=0.6\textwidth]{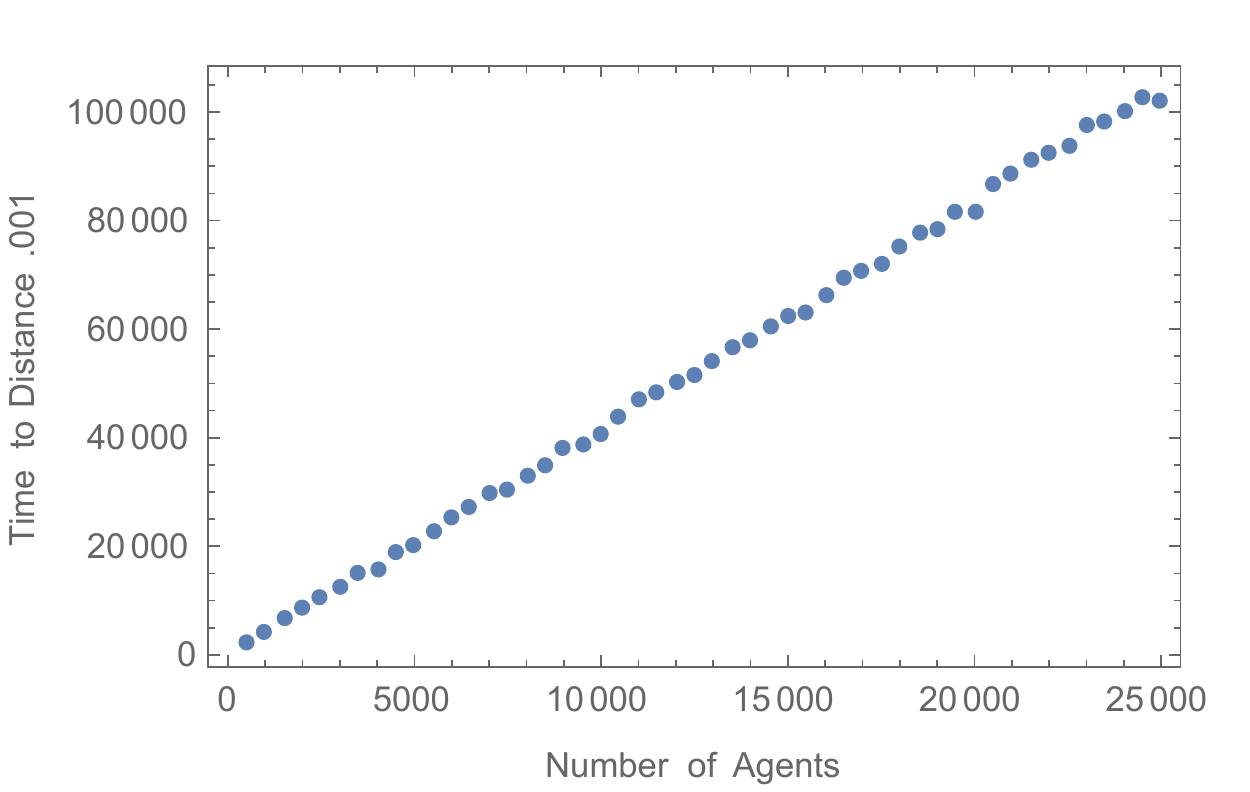}
\caption{Convergence time to the near the steady-state distribution is a constant number of rounds per agent.}
\label{fig:number}
\end{figure}

\begin{figure}[htb]
\centering
\includegraphics[width=0.6\textwidth]{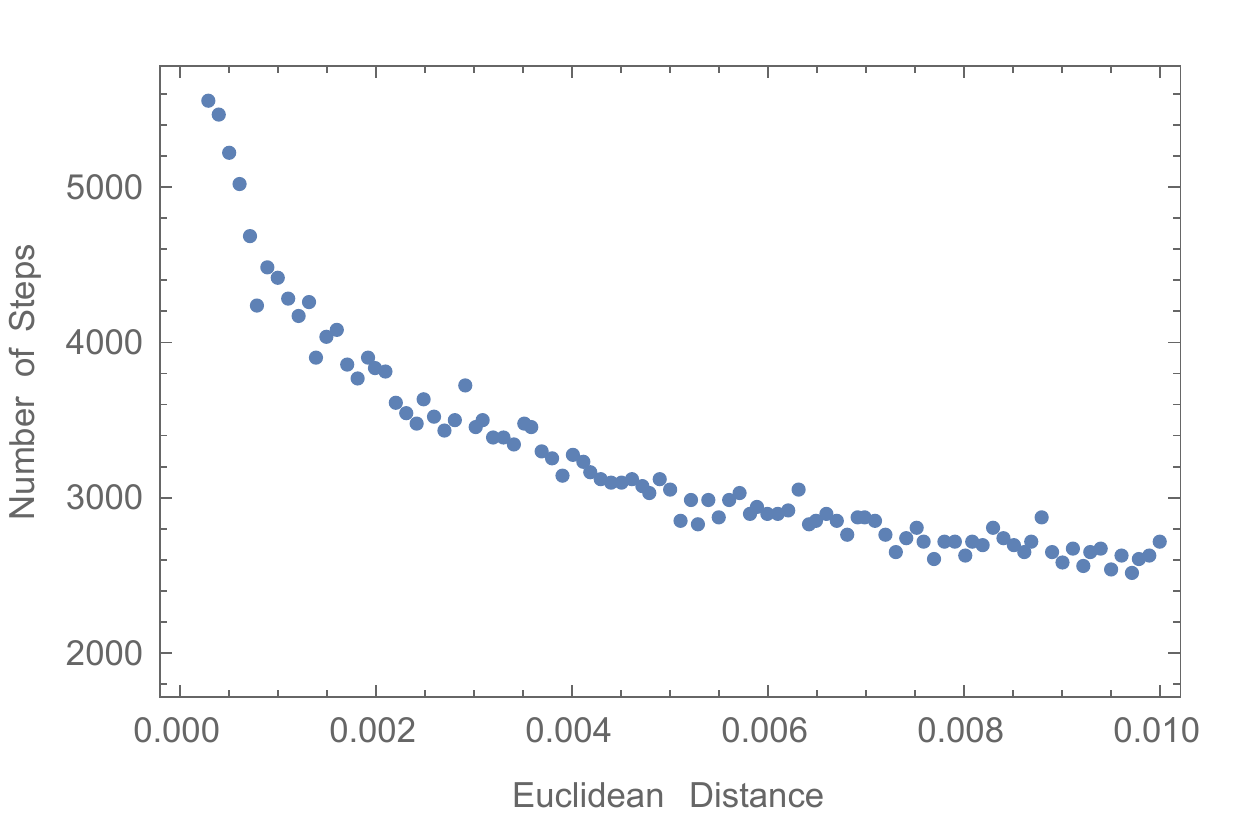}
\caption{Convergence is fast even for very close distances.}
\label{fig:distance}
\end{figure}

We simulated starting with the 
most extreme distribution possible (every agent has either 0 or $k +
1/b -1$ tokens) to 
determine how long it took to get close to the steady-state distribution. 
Figure~\ref{fig:number} shows that even in this unrealistic and extreme case, convergence
takes only a small constant number of rounds per agent. (With these
parameters, 5 rounds per agent suffice.)
Figure~\ref{fig:distance}, which fixes the number of agents at 1000, shows that the
required time does not rise significantly even if we ask for very small distances.

\begin{figure}[htb]
\centering
\includegraphics[width=0.6\textwidth]{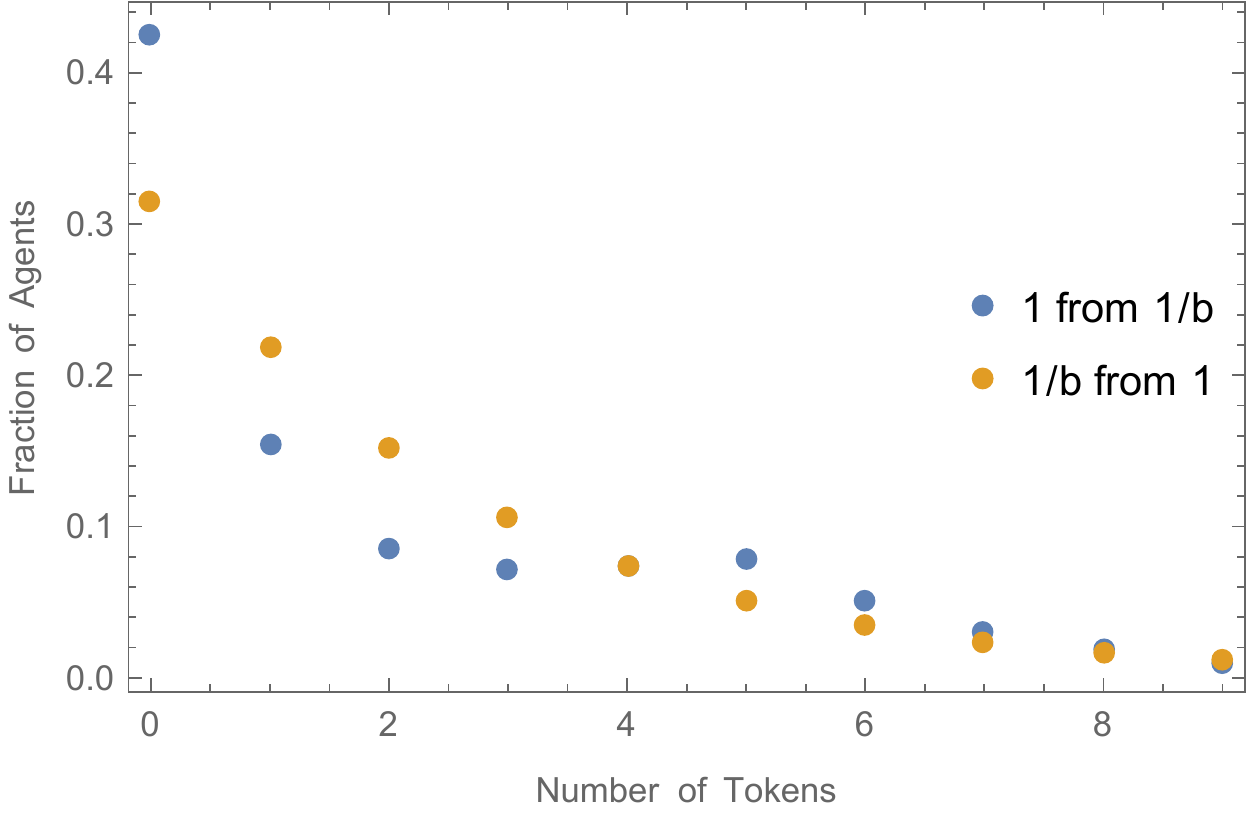}
\caption{Original and Alternate distribution of tokens.}
\label{fig:distributions}
\end{figure}

\begin{figure}[htb]
\centering
\includegraphics[width=0.6\textwidth]{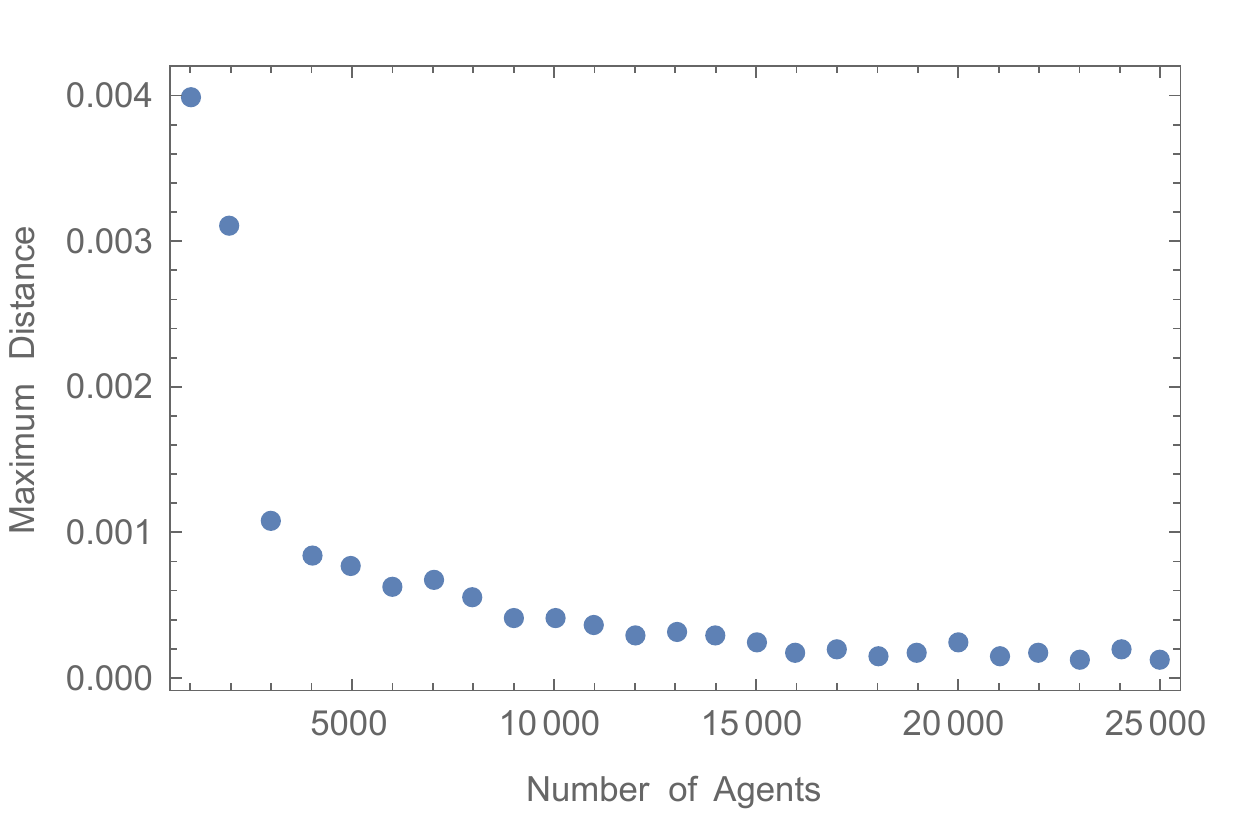}
\caption{The system stays close to the steady-state distribution of tokens.}
\label{fig:close2}
\end{figure}

\begin{figure}[htb]
\centering
\includegraphics[width=0.6\textwidth]{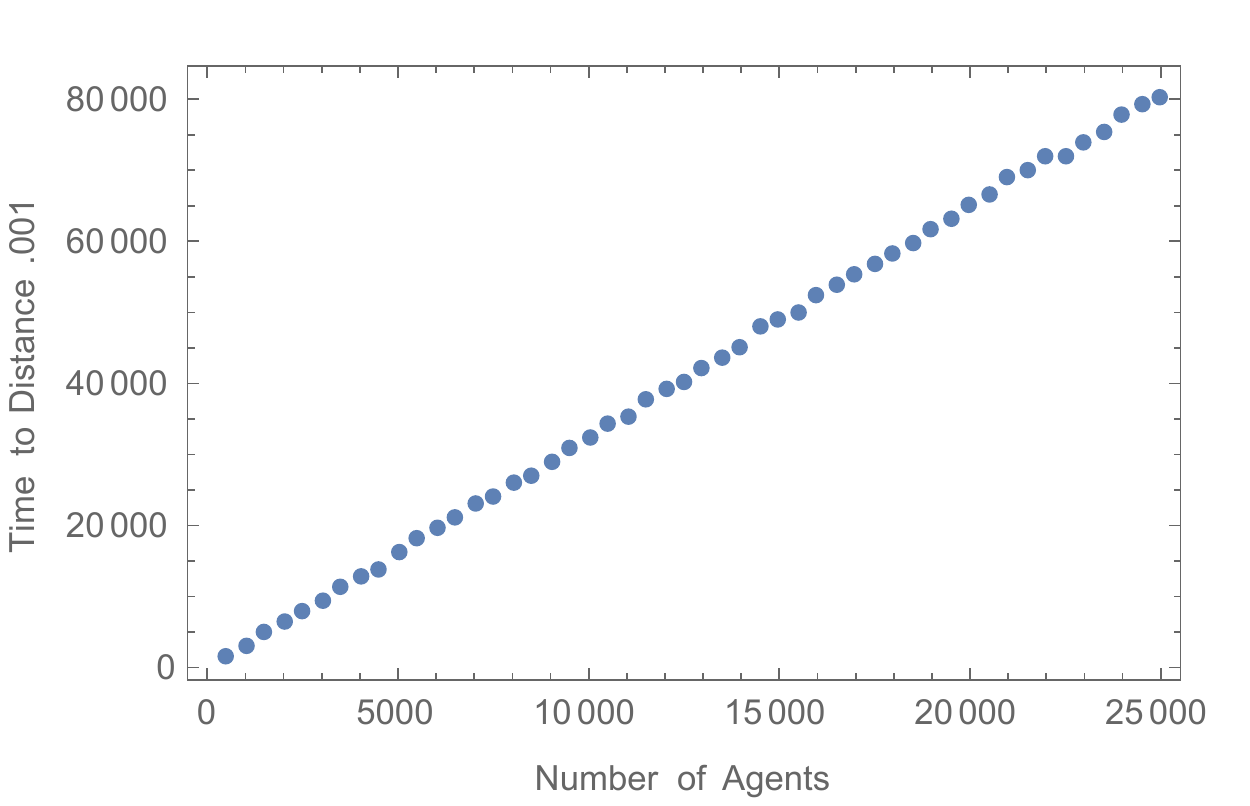}
\caption{Convergence time to the near the steady-state distribution is a constant number of rounds per agent.}
\label{fig:number2}
\end{figure}

\begin{figure}[htb]
\centering
\includegraphics[width=0.6\textwidth]{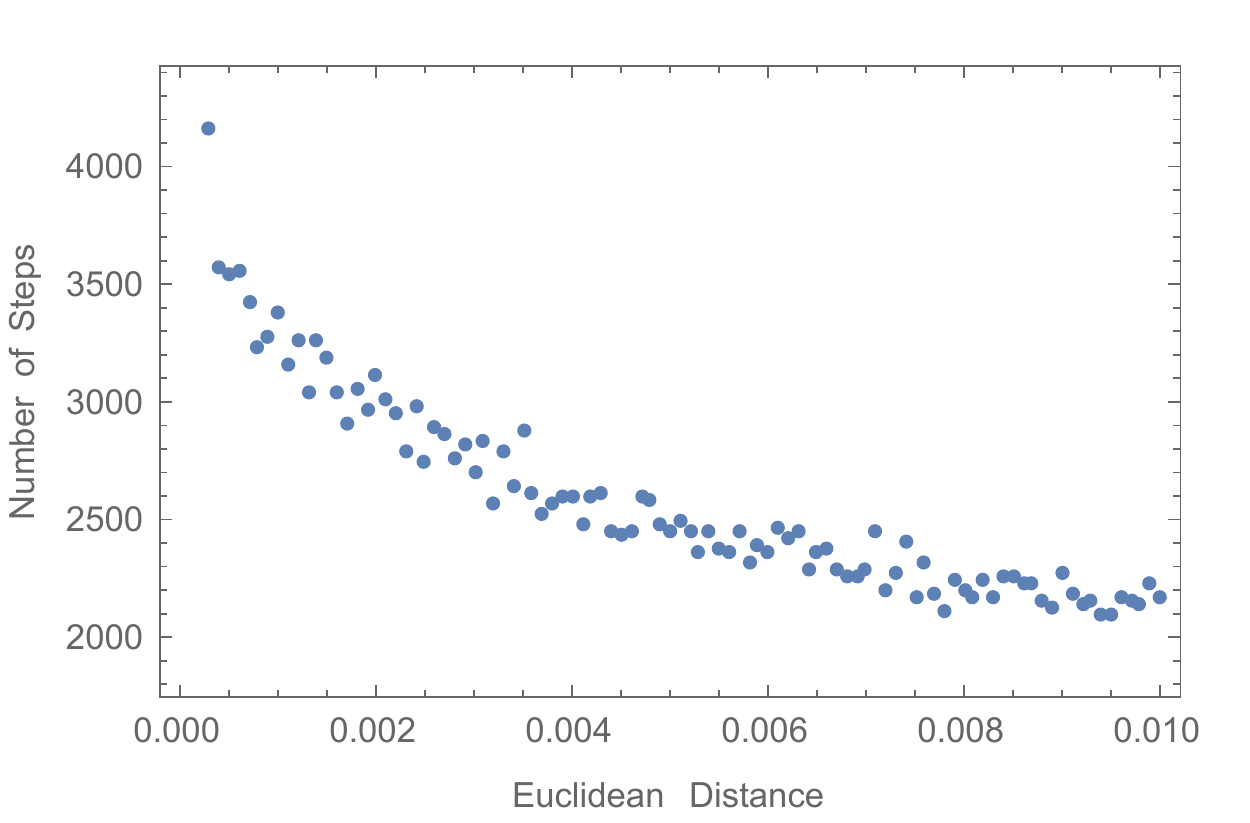}
\caption{Convergence is fast even for very close distances.}
\label{fig:distance2}
\end{figure}

Our model has the somewhat undesirable feature that, when a payment
must be made,
all $1/b$ tokens are taken from a single agent.  It seems more
palatable to take a 
single token from $1/b$ agents instead.  However, this invalidates the
technique used to
calculate the steady-state distribution of wealth.  While other work
has shown that 
it is possible to find other ways of calculating the steady-state
distribution in other settings where the analogous assumption is not
made~\cite{Humbert//:11a}, this  
requires significant effort.  Our intuition strongly suggested that charging
$1/b$ agents one token each should have no effect on the results,
although we could not find a formal proof.  
Therefore, we decided to empirically validate the existence of a
steady-state distribution of tokens.  To do so, we ran this alternate
mechanism for 
100 million steps, sampling the distribution every 20000 steps, and
took the sample 
mean as our estimate of the true steady-state distribution.
This change affected the final steady-state
distribution more than we expected. 
The results are shown in Figures \ref{fig:distributions}--\ref{fig:distance2}.

The blue dots in 
Figure~\ref{fig:distributions} show the new 
steady-state distribution;
the original distribution is described by the orange dots.
The distance between these distributions
is $0.152$, 
which is two orders of magnitude larger than the variation around the
steady-state 
distribution that we saw Figure~\ref{fig:close}.

We verified that this is in fact the steady-state distribution by rerunning all three
simulations and calculating distances from it.  The results, shown in
Figures~\ref{fig:close2}--\ref{fig:distance2}, are essentially the same.  The results
of this change seem 
positive overall: agents do not face sudden large drops in their
supply of tokens and convergence is, if anything, modestly faster.
There are more 
agents without tokens, but this could be mitigated by using a larger number of
tokens.

\section{Related Work}

Our analysis of the behaviour and incentives of the token economy draws
heavily on prior work on scrip systems
by Kash et al.~\citeyear{Friedman//:06a,Kash//:12a}. We
adopt many of their techniques, but extend their analysis to a variant model
that applies to our setting.  Other work has shown that changing the random
volunteer procedure can improve welfare~\cite{Johnson//:14a} and that this
approach still works if more than one agent must be hired to perform
work~\cite{Humbert//:11a}.  Work from the systems community has
looked at practical details such as
the efficient implementation of a token
bank~\cite{Vishnumurthy//:03a}.

Another strand of related work is on game-theoretic models of norm emergence.
Axelrod \citeyear{Axelrod:86a} showed by means of simulations 
how norms could emerge given simple game rules where players
punish each other for violations (and punish players who don't punish
violations), and a
number of norm enforcement mechanisms with good incentive properties 
have been analysed ~\cite{Kandori//:92a,Ellison//:94a}.
Axelrod's work has been extended by Mahmoud et al.\  \citeyear{Mahmoud//:12a} to general scenarios and to incorporate learning.
Da Pinnick et al.\ \citeyear{Pinninck//:10a} proposed
a distributed norm enforcement mechanism 
that uses ostracism as punishment, and showed both
analytically and experimentally that it provides an upper bound on the
number of norm violations.

There is a significant amount of work in the MAS literature on infrastructures for implementing normative organisations, monitoring for norm violations, and compensating violations through sanctions. 
One common approach involves the use of additional components or agents to implement the normative organisation. For example, Boella and van der Torre \citeyear{Boella/vdTorre:03a} propose `defender agents' which detect and punish norm violations. Esteva et al.\ \citeyear{Esteva//:04a} propose the use of `governors' to monitor and regiment message exchanges between agents; each agent is associated with a governor, and all interactions with other agents are filtered by the governor to ensure compliance with norms.
Grizard et al.\ \citeyear{Grizard//:07a}, propose an approach in which a separate system of `controller agents'  monitor norm violations and apply reputational sanctions to `application agents' in a MAS; application agents avoid interactions with other application agents that have low reputation, hence eventually excluding bad agents from the system. 
Modgil et al.\ \citeyear{Modgil//:09a} propose a two-layer approach, in which `trusted observers' relay observations of states of interest referenced by norms to `monitor agents' responsible for determining whether a norm has been violated
(a similar approach is described by Criado et al. \cite{Criado//:11a}).
H\"ubner et al.\ \citeyear{Hubner//:10b} describe an approach in which `organizational agents' monitor interactions between agents mediated by `organizational artifacts'.
Balke et al.\ \citeyear{Balke//:13a} have used simulation to investigate the effectiveness and costs of paying `enforcement agents' to monitor norm violations in a wireless mobile grid scenario; the mechanism they propose for rewarding enforcement agents results in a cost to the MAS (in their setting, the telecommunications company), and they assume sanction-based enforcement (agents who violate the norm are punished by the telecommunications company). 
Testerink et al.\ \citeyear{Testerink//:14a} consider the problem of monitoring and enforcement by a network of normative organisations in which each normative organisation has only partial information about the actions of the agents and is capable of only local enforcement (by sanctioning).

These approaches are distributed, but the responsibility for
monitoring still lies with the normative organisation, and the cost of
monitoring is borne by the MAS,  
either in the cost of running additional system components which
monitor and regulate interactions (\eg
\cite{Boella/vdTorre:03a,Esteva//:04a}) or by paying some agents to
monitor the rest (\eg \cite{Balke//:13a}).  
Fagundes et al.\ \citeyear{Fagundes//:14a} have explored the tradeoff
between the efficiency and cost of norm enforcement in stochastic
environments, to identify scenarios in which monitoring can be funded
by sanctions levied on violating agents while at the same time keeping
the number of violations within a tolerable level. 
However, in an open multi-agent system, approaches in which norm
enforcement is based on sanctioning (\eg
\cite{Grizard//:07a,Testerink//:14a}) may be susceptible to sybil
attacks; sanctioned agents may simply leave the system and rejoin
under a different id.  
In contrast, in our approach, the cost of monitoring is borne by the agents using the MAS. Moreover, agents cannot benefit by dropping out and rejoining the system, and monitoring is m-resilient against collusion by monitoring agents.



\section{Conclusion}

We propose an approach to norm monitoring and show that, for sufficiently large MAS, 
perfect monitoring (and hence enforcement) can be achieved when violations are inadvertent. When violations are strategic, the probability of a norm violation can be made arbitrarily small. This is achieved at no cost to the MAS and without
assuming that fines can be used to pay for monitoring. Instead, we achieve
perfect or near perfect enforcement using techniques adapted from scrip 
systems \cite{Friedman//:06a}.
Our approach is limited to monitoring norms which forbid single actions (or resulting states). In contrast, approaches such as \cite{Esteva//:04a,Modgil//:09a,Hubner//:10b} are capable of monitoring conditional norms which specify complex behaviours, such as multi-step protocols. 
We leave extending our approach to such conditional norms to future work.

\fullv{
\section{Appendix: Sketch of Proof of Theorems~\ref{thm1} and ~\ref{thm2}}\label{sec:proof}
In the setting of Friedman et al. \citeyear{Friedman//:06a} (denoted
FHK in this section), there is someone who wants work done and an
agent who is willing to do it.  The agent who wants work done gives a token to
the agent willing to do it (chosen among volunteers, just as in our
setting).  In our setting, there is an agent who wants to post
something; it plays the same role as the agent who wants work done in
the FHK system.  But now the posting agent gives a token to a random
agent, and the agent performing work (monitoring) gets paid only if
he detects a problem.  Moreover, it is not paid by the agent doing the
posting, but by a random agent; and it is not paid one token, but
$1/b$ (or $1/\beta^*$) tokens.  While these seem to be significant
differences, the argument used by FHK to prove that there exists an
equilibrium in threshold strategies goes through almost without
change.

We briefly sketch the key features of the FHK argument here, and the
differences in our setting.
Suppose that all players are following a threshold strategy with
threshold $k$.  What is the best response for a given agent?  
In particular, when should the agent volunteer?

Clearly, the one thing that the agent is concerned about is that it
will run out of  
tokens before it is next chosen to monitor, and thus not be able to post when 
it has the opportunity to do so.
The likelihood of this
happening depends on how many other volunteers there are each time the
agent volunteers.  To take an extreme case, if the agent can be sure
that it will be the only volunteer, then it is safe waiting until it
has one token left.  It is unlikely that it will get a chance to make a
posting twice before it is able to earn a token.  On the other hand,
if there will be lots of competition each time it volunteers, then it
would be better to use a higher threshold.

So the first step in the proof is to get an accurate estimate of how
many volunteers there will be at each step.  To do this, we view the system as
a Markov chain, where the state of the system is characterised by how
many tokens each agent has.  For simplicity, we take the Markov chain
to consist of all states reachable from some fixed initial state,
assuming that all players follow a fixed threshold-$k$ strategy and
all players have an integral number of tokens, bounded by $k$.
In each state, it is straightforward to
compute the probability of a transition to another state: each agent
will be chosen to make a posting with probability $1/n$; we know
whether that agent has a token (and thus can post something); we
know exactly who will volunteer to monitor (all agents with less than
$k$ tokens); and, since a volunteer is chosen at random among the
volunteers, we also know the likelihood that an agent will be chosen
to monitor.  The key observation is that, in the FHK setting, this
Markov chain is \emph{reversible}: the 
probability of a transition from a state $s$ to a
state $s' \ne s$ is the same as the probability of the reverse transition
from $s'$ to $s$. 

In the FHK setting, reversibility is easy to check: the only
such transition that can occur is from a state
where an agent $i$ performs work for an agent $j$.  
Then in state $s'$, $j$ has one fewer token than in state $s$, and $i$
has one more.   
This transition occurs
with probability $1/nm$, where $m$ is the number of agents in state
$s$ other than $j$ who have fewer than $k$ tokens: 
$i$ is chosen in state $s$ with probability $1/m$, and 
$j$ is chosen with independent probability $1/n$.  
The transition from $s'$
to $s$ also has probability $1/nm$.  
Agent $i$ is chosen in $s'$
with probability $1/n$ (and 
$i$ is guaranteed to have at least one
token in $s'$, since 
it received the token from $j$) and there are again $m$
volunteers: all the ones that volunteered in state $s$ other than $i$
together with agent $j$ (who must have less than $k$
tokens in state $s'$, since $j$ gave a token to $i$).

In the inadvertent setting, we again take the Markov chain to consist of all
states reachable with positive probability from some fixed starting
state, with everyone following the threshold strategy.
Reversibility still holds,
but we have 
to work a little harder to show it.  For simplicity, we assume that
the act of the person wanting to do monitoring paying one token to a
random agent with less the maximum number of tokens is distinct from
the act where an agent who discovers a violation is paid $1/b$ tokens 
by a random agent with at least $1/b$ tokens.  In the first
transition, the agent $i$ who makes a post ends up with one less token
and the random agent $j$ who gets it has one more token.  This
transition happens with probability $1/nm$, where $m$ is the number of
agents other than $i$ who have less than $k+1/b$ tokens.
The reverse transition happens with the same probability: $j$ is
chosen to post something with probability $1/n$, $j$ can make a
posting since it received a token from $i$, and the agents who have
less than $k+1/b$ tokens other than $j$ is still $m$:
all the agents who had less than $k+1/b$ tokens in state
$s$, together with $i$ (who must have less than $k+1/b$
tokens, since it gave a token to $j$).

Now consider the transition from $s$ to $s'$ where $i$ gives $j$ $1/b$
tokens because $j$ discovered a violation.  This transition occurs
with probability $b/mm'$, where $m$ is the number of agents other than
$i$ with less than $k$ tokens in $s$ (these are the volunteers) and
$m'$ is the number of agents with at least $1/b$ tokens.  
Again the reverse transition happens with the same probability: now
$j$ must have at least $1/b$ tokens and $i$ must have less than $k$
tokens, so $i$ is a potential volunteer and $j$ is 
an agent who can pay $i$.  

The rest of the argument now continues as in \cite{Friedman//:06a}, so
we just sketch the details.
Clearly the Markov chain is finite, given our assumption that $b$ is
rational: we can easily bound the number of possible states, since $b$ is
rational and players have at most $k+1/b$ tokens.%
\footnote{Here is where we need the technical assumption that $b$ is rational.}
It immediately follows from reversibility and our
assumptions that the states in the Markov chain consist of all states
reachable from some initial starting state that 
Markov chain is \emph{irreducible}: every state is reachable from
every other state.  Finally, the same arguments as in FHK show that
the Markov chain is \emph{aperiodic}: for every state $s$, there exist
to cycles from $s$ to itself such that the gcd of their lengths is 1.

It is well known \cite{Resnick} that every finite, reversible,
aperiodic, and irreducible Markov chain has a \emph{limit
distribution} $\pi$ (where $\pi(s)$ is the fraction of the time that
the Markov chain is in state $s$) and in this limit distribution,  
all states are equally likely.  
However, we are not so interested in the probability of a given state; we
are interested in the probability of the distribution of tokens; in
particular we are interested in the fraction of agents that have less
than the threshold number of tokens, because this will tell us how
many volunteers there will be.  To understand the difference, consider
a system with two tokens.  There are $n$ states where agent has both the
tokens, and 
$n \choose 2$ states where two agents have one token
each.  So although all states are equally likely, the distribution 
where  $2/n$ of the agents have one token and the rest have none is
far more likely than the distribution  where $1/n$ of the agents have two 
tokens, and the rest have none.

Using standard techniques, it can be
shown that the distribution that maximises entropy is overwhelmingly
more likely than the rest.  So, as long as $n$ is sufficiently large, the
number of agents with less than the  threshold $k$ of tokens is, with
extremely high probability, $\gamma n$, where $\gamma$ is the
probability of having less than $k$ tokens according to the maximum
entropy distribution.  Once the agent knows how much competition she
will face, it is easy to compute a best response, and to show that
there is a best response in threshold strategies.

\newcommand{\BR}{\mathit{BR}}
Let $\BR(k)$ denote the agent's best-response threshold strategy if
all other agents are using threshold $k$.  FHK show that $\BR(k)$ is
monotonically increasing in $k$; moreover, for $\delta$ sufficiently
close to 1, there exists $k$ such that $\BR(k) > k$.  It follows using
standard arguments that $\BR$ has a fixed point that is greater than
$k$. That fixed-point is an $\epsilon$-Nash equilibrium, for if
$\BR(k) = k$ and all agents are playing a threshold strategy, then
playing a threshold of $k$ is an $\epsilon$-best response.  (The
$\epsilon$ accounts for the fact that the the distribution of money is
not exactly described by the maximum entropy distribution, and even
with the maximum entropy distribution, there is a very small chance
that playing $k$ is not a best response.)   This completes the proof
sketch of Theorem~\ref{thm1}. 

The argument in the case of Theorem~\ref{thm2} is quite similar.  Here
the setting appears more complicated, because agents have two
strategic choices: deciding whether or not to post bad material, and
deciding whether or not to volunteer.  However, as shown earlier, our choice of
parameters guarantees that an agent's payoff is independent of its
strategy regarding whether or not to post bad material.  Indeed, if we
fix agent $i$'s strategy for deciding whether or not to volunteer, and
fix the strategies of all other agents, all choices
of strategy for deciding whether or not to post bad material, even
ones that are history-dependent and correlated the agent $i$'s
threshold and the strategies of all other agents, give $i$ the same utility.
So, we can assume without loss of generality that in equilibrium,
agent $i$ makes a bad posting with probability $\beta^*$.

Now essentially the same argument as that used in the proof of
Theorem~\ref{thm1} shows that there is an equilibrium where agents'
make their choices regarding volunteering according to some threshold
strategy.  In particular, the reversibility argument still holds; the
probability just need to be multiplied by $1-1/\kappa$.
}


\newpage

\bibliographystyle{named}
\bibliography{norms}  

\end{document}